%% file: lstar_prd2011.tex
\documentclass[twocolumn,showpacs,preprintnumbers,amsmath,amssymb]{revtex4} 

\usepackage{graphicx}
\usepackage{dcolumn}
\usepackage{multirow}

\usepackage{atlasphysics} 
\usepackage[hyperindex,breaklinks]{hyperref} 

\def\lstar{\ensuremath{{\ell^{*}}}}

\def\zpjet{\ensuremath{Z +{\rm jets}}}
\def\zpgam{\ensuremath{Z + \gamma}}
\def\wpjet{\ensuremath{W+{\rm jets}}}

\def\ptmu{\ensuremath{\pt^{\mu}}}
\def\pte{\ensuremath{\pt^{\rm e}}} 
\def\ptlepton{\ensuremath{\pt^{\ell}}} 
\def\ptgam{\ensuremath{\pt^{\gamma}}} 

\def\ifb{\mbox{fb$^{-1}$}}

\def\mcatnlo{MC@NLO}
\def\pythia{{\sc Pythia}}
\def\sherpa{{\sc Sherpa}}
\def\herwig{{\sc Herwig}}
\def\jimmy{{\sc Jimmy}}
\def\alpgen{{\sc Alpgen}}
\def\comphep{{\sc CompHEP}}

\def\geant{{\sc Geant}}
\def\photos{{\sc photos}}
\def\medium{{\it medium}}
\def\tight{{\it tight}}

\begin{document}

\preprint{CERN-PH-EP-2011-224}
\preprint{Submitted to Physical Review D}
\vspace*{0.1cm}
\title{Search for excited leptons in proton--proton collisions \\ at $\sqrt{s}=7$~TeV with the ATLAS detector}
\author{The ATLAS Collaboration}

\begin{abstract}  
The ATLAS detector is used to search for excited leptons in the electromagnetic radiative decay channel 
$\ell^*\rightarrow \ell\gamma$. Results are presented based on the analysis of $pp$ collisions at a 
center-of-mass energy of 7~TeV corresponding to an integrated luminosity of 2.05~\ifb. No evidence for 
excited leptons is found, and limits are set on the compositeness scale $\Lambda$ as a function of the 
excited lepton mass $m_{\lstar}$. In the special case where $\Lambda = m_{\lstar}$, excited electron and 
muon masses below 1.87~TeV and 1.75~TeV are excluded at 95\% C.L., respectively.
\end{abstract}


\maketitle

\section{Introduction}

The Standard Model (SM) of particle physics is an extremely successful effective theory which has been extensively 
tested over the past forty years. However, a number of fundamental questions are left unanswered.  In particular, the 
SM does not provide an explanation for the source of the mass hierarchy and the generational structure of quarks 
and leptons.  Compositeness models address these questions by proposing that quarks and leptons are composed of 
hypothetical constituents named preons \cite{patisalam}. In these models, quarks and leptons are the lowest-energy 
bound states of these hypothetical particles.  New interactions among quarks and leptons should then be visible at 
the scale of the constituents' binding energies, and give rise to excited states.  At the Large Hadron Collider 
(LHC), excited lepton $\lstar$ production via four-fermion contact interactions can be described by the effective 
Lagrangian \cite{BaurSpiraZerwas}
\begin{displaymath} 
{\cal L}_{\rm contact} =\frac{g_\ast^2}{2\Lambda^2} j^\mu j_\mu, 
\end{displaymath} 
\noindent
where $g_\ast^2$ is the coupling constant, $\Lambda$ is the compositeness scale, and $j_\mu$ is the 
fermion current 
\begin{displaymath} 
j_\mu = \eta_L\overline{f}_L\gamma_\mu f_L + \eta'_L \overline{f}_L^\ast\gamma_\mu f_L^\ast + 
\eta_L''\overline{f}_L^\ast\gamma_\mu f_L + h.c. + (L\rightarrow R). 
\end{displaymath} 
\noindent For simplicity and consistency with recent searches, the following prescription is used: $g_\ast^2=4\pi$, 
$\eta_L = \eta'_L=\eta''_L=1$, and $\eta_R = \eta'_R=\eta''_R=0$ such that chiral symmetry is conserved 
\cite{eichten}\cite{contactI}. The above ansatz ignores underlying preon dynamics and is valid as long as the mass of the 
excited leptons is below the scale $\Lambda$.  In the well-studied case of the homodoublet-type 
\lstar~\cite{hagiwara,cabbibo,BaurSpiraZerwas}, the relevant gauge-mediated Lagrangian describing transitions between 
excited and ground-state leptons is
\begin{displaymath}
{\cal L}_{\rm GM} = \frac{1}{2\Lambda} \overline{\ell}_R^\ast \sigma^{\mu\nu}\left[g f 
\frac{\tau^a}{2}W_{\mu\nu}^a+g'f'\frac{Y}{2}B_{\mu\nu} \right] \ell_L + h.c.,
\label{eqn:lagrangian_trans}
\end{displaymath}
\noindent
where $\ell_L$ is the lepton field, $W_{\mu\nu}$ and $B_{\mu\nu}$ are the $SU(2)_L$ and $U(1)_Y$ field strength tensors, 
$g$ and $g'$ are the respective electroweak couplings, and $f$~and $f'$ are phenomenological constants chosen to be equal to 
1.  The ${\cal L}_{\rm GM} $ term allows the decay of excited leptons via the electromagnetic radiative mode 
${\ell^\ast}^\pm \to \ell^\pm \gamma$, a very clean signature which is exploited in this search.  For a fixed value 
of $\Lambda$, the branching ratio $B({\ell^\ast}^\pm \to \ell^\pm \gamma)$ decreases rapidly with increasing 
$\lstar$ mass.  For $\Lambda = 2$~\tev, $B({\ell^\ast}^\pm \to \ell^\pm \gamma)$ is 30\% for $m_\lstar = 0.2$~\tev\ 
and decreases exponentially to about 2.3\% for $m_\lstar = 2$~\tev.

Previous searches at LEP \cite{LEPresults}, HERA \cite{HERAresults}, and the Tevatron \cite{TEVATRONresults} have found no 
evidence for such excited leptons.  For the case where $\Lambda = m_{\lstar}$, the CMS experiment has excluded masses 
below 1.07~\tev\ for $e^*$ and 1.09~\tev\ for $\mu^*$ at the 95\% credibility level (C.L.)
\cite{Chatrchyan:2011pg}.  


\section{Analysis strategy}

This article reports on searches for excited electrons and muons in the $\ell^*\rightarrow \ell \gamma$ channel 
based on 2.05~\ifb\ of 7~TeV $pp$ collision data recorded in 2011 with the ATLAS 
detector~\cite{atlas:detector}. The benchmark signal model considered is based upon theoretical 
calculations from Ref.~\cite{BaurSpiraZerwas}. In this model, excited leptons may be produced singly 
via $q\overline{q}\rightarrow \ell^*\overline{\ell}$ or in pairs via $q\overline{q}\rightarrow 
\ell^*\overline{\ell}^*$, due to contact interactions.  As the cross section for pair production is 
much less than for single production, the search for excited leptons is based on the search for 
events with $\ell\overline{\ell}\gamma$ in the final state: three very energetic particles, isolated, 
and well separated from one another.

For both the $e^*$ and $\mu^*$ searches, the dominant background arises from Drell-Yan (DY) processes 
accompanied by either a prompt photon from initial or final state radiation (\zpgam) or by a jet 
misidentified as a photon (\zpjet). The dominant irreducible \zpgam\ background results in the same 
final state as the signal, whereas \zpjet\ background can be highly suppressed by imposing stringent 
requirements on the quality of the reconstructed photon candidate. Small contributions from \ttbar\ 
and diboson ($WW$, $WZ$ and $ZZ$) production are also present in both channels. \wpjet\ events, as 
well as semileptonic decays of heavy flavor hadrons, and multijet events can be heavily suppressed by 
requiring the leptons and photons to be isolated and thus have a negligible contribution to the total 
background.

The signature for excited leptons can present itself as a peak in the invariant mass of the $\ell + 
\gamma$ system because the width of the $\ell^*$ is predicted to be narrower than the detector mass 
resolution for excited lepton masses $m_{\lstar} \lesssim 0.5 \Lambda$.  This peak could be easily 
resolved from the \zpgam\ background.  However, it is difficult to identify which of the two leading 
leptons in the event comes from the \lstar\ decay. To avoid this ambiguity, one can search for an 
excess in the $\ell\overline{\ell}\gamma$ invariant mass ($m_{\ell\ell\gamma}$) spectrum.  This 
approach is effective for the whole $m_{\lstar} - \Lambda$ parameter space probed as one can search 
for an excess of events with $m_{\ell\ell\gamma} > 350$~\gev, which defines a nearly background-free 
signal region.
Optimization studies demonstrate that the observable $m_{\ell\ell\gamma}$ provides better
signal sensitivity than $m_{\ell\gamma}$, particularly for lower \lstar\ masses.
The analysis strategy therefore relies on $m_{\ell\ell\gamma}$ for the statistical interpretation of the results.

\section{ATLAS detector}

ATLAS is a multi-purpose detector with a forward-backward symmetric cylindrical geometry and nearly 
4$\pi$ coverage in solid angle. It consists of an inner tracking detector immersed in a 2~T 
solenoidal field, electromagnetic and hadronic calorimeters, and a muon spectrometer. Charged 
particle tracks and vertices are reconstructed in silicon-based pixel and microstrip tracking 
detectors that cover $|\eta| <$ 2.5 and transition radiation detectors extending to $|\eta| <$ 2.0 
\cite{coordinates}. A hermetic calorimetry system, which covers $|\eta|$ $<$ 4.9, surrounds the 
superconducting solenoid. The liquid-argon electromagnetic calorimeter, which plays an important role 
in electron and photon identification and measurement, is finely segmented.  It has a readout 
granularity varying by layer and cells as small as $0.025 \times 0.025$ in $\eta \times \phi$, and 
extends to $|\eta| < 2.5$ to provide excellent energy and position resolution. Hadron calorimetry is 
provided by an iron-scintillator tile calorimeter in the central rapidity range $|\eta| < 1.7$ and a 
liquid-argon calorimeter with copper and tungsten as absorber material in the rapidity range $1.5 < 
|\eta| < 4.9$. Outside the calorimeter, there is a muon spectrometer which is designed to identify 
muons and measure their momenta with high precision. The muon spectrometer comprises three toroidal 
air-core magnet systems: one for the barrel and one per endcap, each composed of eight coils.  Three 
layers of drift tube chambers and/or cathode strip chambers provide precision ($\eta$) coordinates 
for momentum measurement in the region $|\eta |< 2.7$.  A muon trigger system consisting of resistive 
plate chambers in the barrel and thin-gap chambers for $|\eta| > 1$ provides triggering capability up 
to $|\eta| = 2.4$ and measurements of the $\phi$ coordinate.

\section{Simulated samples}

The excited lepton signal samples are generated based on calculations from 
Ref.~\cite{BaurSpiraZerwas} at leading order (LO) with \comphep\ 4.5.1 \cite{comphep} interfaced with 
\pythia\ 6.421 to handle parton showers and hadronization \cite{CompHep_Pythia,pythia}, using 
MRST2007 LO* \cite{mrst} parton distribution functions (PDFs).  Only single production of excited 
leptons is simulated, with the \lstar\ decaying exclusively via the electromagnetic channel. The 
\zpgam\ sample is generated with \sherpa\ 1.2.3 \cite{SherpaManual} using CTEQ6.6 \cite{cteq} PDFs, 
requiring the dilepton mass to be above 40~\gev. To avoid phase-space regions where matrix elements 
diverge, the angular separation between the photon and leptons is required to be $R(\ell,\gamma) = 
\sqrt{(\Delta \eta)^2 + (\Delta \phi)^2}> 0.5$ and the transverse momentum (\pt) of the photon is 
required to be $\ptgam>10$~GeV.  To ensure adequate statistics at large $m_{\ell\ell\gamma}$, an 
additional \zpgam\ sample is generated with $\ptgam>40$~GeV, and is equivalent to $\sim 300$~\ifb\ of 
data.  The \zpjet\ background is generated with \alpgen\ 2.13 \cite{Mangano:2002ea}, while the 
\ttbar\ background is produced with \mcatnlo\ 3.41 \cite{mcatnlo}.  In both cases, \jimmy\ 4.31 
\cite{jimmy} is used to describe multiple parton interactions and \herwig\ 6.510 \cite{herwig} is 
used to simulate the remaining underlying event and parton showers and hadronization.  CTEQ6.6 PDFs 
are used for both backgrounds.  To remove overlaps between the \zpjet\ and the \zpgam\ samples, 
\zpjet\ events with prompt energetic photons are rejected if the photon-lepton separation is such 
that $R(\ell,\gamma) > 0.5$.  The diboson processes are generated with \herwig\ using MRST2007 LO* 
PDFs. For all samples, final-state photon radiation is handled via \photos\ \cite{fsr_ref}.  The 
generated samples are then processed through a detailed detector simulation \cite{atlas:sim} based on 
\geant4 \cite{geant} to propagate the particles and account for the detector response.  A large 
sample of MC minimum bias events is then mixed with the signal and background MC events to simulate 
pile-up from additional $pp$ collisions.  Simulations are normalized on an event-by-event basis such 
that the distribution of the number of interactions per event agrees with the spectrum observed in 
data.

Although \sherpa\ includes higher-order QCD contributions beyond the \zpgam\ Born amplitude, such as 
the real emission of partons in the initial state, it omits virtual corrections. For this reason, the 
\zpgam\ cross section is calculated at next-to-leading order ($\sigma_{\rm NLO}$) using MCFM 
\cite{MCFM} with MSTW 2008 NLO PDFs \cite{mstw}.  The theoretical precision of the $\sigma_{\rm NLO}$ 
estimate is $\sim 6\%$, and the ratio $\sigma_{\rm NLO} / \sigma_{\rm SHERPA}$ is used to determine a 
correction factor as a function of $m_{\ell\ell\gamma}$. The \zpjet\ cross section is initially 
normalized to predictions calculated at next-to-next-to-leading order (NNLO) in perturbative QCD as 
determined by the FEWZ \cite{fewz} program using MSTW 2008 NNLO PDFs. Since the misidentification of 
jets as photons is not well modeled, the \zpjet\ prediction is adjusted at the analysis level using 
data-driven techniques described below.  Cross sections for diboson processes are known at NLO with 
an uncertainty of 5\%, while the \ttbar\ cross section is predicted at approximate-NNLO, with better 
than 10\% uncertainty \cite{Moch:2008qy,Langenfeld:2009tc}.

\section{Data and preselection}

The data, which correspond to a total integrated luminosity of 2.05 \ifb, were collected in 2011 
during stable beam periods of 7~TeV $pp$ collisions. For the $e^*$ search, events are required to 
pass the lowest unprescaled single electron trigger available.  For the first half of the data this 
corresponds to a \pte\ threshold of 20~\gev, and a \pte\ threshold of 22~\gev\ for the later runs.  
For the $\mu^*$ search, a single muon trigger with matching tracks in the muon spectrometer and inner 
detector with combined $\ptmu > 22$~\gev\ is used to select events. In addition, events with a muon 
with $\ptmu>40$~\gev\ in the muon spectrometer are also kept. Collision candidates are then 
identified by requiring a primary vertex with a $z$ position along the beam line of $|z|<200$~mm and
at least three associated charged particle tracks with $\pt > 0.4$~\gev.

The lepton selection consists of the same requirements used in the ATLAS search 
for new heavy resonances decaying to dileptons \cite{Zprime2011}. Electron candidates are formed from clusters of cells in 
the electromagnetic calorimeter associated with a charged particle track in the inner detector. For 
the $e^*$ search, two electron candidates with $\pte > 25$~\gev\ and $|\eta| < 2.47$ are
required. Electrons within the transition region $1.37 < |\eta| < 1.52$ between the barrel and the 
endcap calorimeters are rejected.  The \medium\ electron identification criteria \cite{egamma_id} on 
the transverse shower shape, the longitudinal leakage into the hadronic calorimeter, and the 
association with an inner detector track are applied to the cluster.  The electron's reconstructed 
energy is obtained from the calorimeter measurement and its direction from the associated track. A 
hit in the first active pixel layer is required to suppress background from photon conversions. To 
further suppress background from jets, the leading electron is required to be isolated by demanding 
that the sum of the transverse energies in the cells around the electron direction in a cone of 
radius $R<0.2$ be less than 7~\gev.  The core of the electron energy deposition is excluded and the 
sum is corrected for transverse shower leakage and pile-up from additional $pp$ collisions to make 
the isolation variable essentially independent of \pte\ \cite{ambientenergy}. 
In cases where more 
than two electrons are found to satisfy the above requirements, the pair with the largest invariant 
mass is chosen.  To minimize the impact of possible charge misidentification, the electrons are not 
required to have opposite electric charges.

Muon tracks are reconstructed independently in both the inner detector and muon 
spectrometer, and their momenta are determined from a combined fit to these two 
measurements. For the $\mu^*$ search, two muons with $\ptmu > 25$~\gev\ are required. 
To optimize the momentum resolution, each muon candidate is required to have a 
minimum number of hits in the inner detector and to have at least three hits in each 
of the inner, middle, and outer layers of the muon spectrometer. This requirement results in a 
muon fiducial acceptance of $|\eta|<2.5$. Muons with hits in the barrel-endcap 
overlap regions of the muon spectrometer are discarded because of large residual 
misalignments.  The effects of misalignments and intrinsic position resolution are 
otherwise included in the simulation. The \ptmu\ resolution at 1~TeV ranges from 13\% 
to 20\%. To suppress background from cosmic rays, the muon tracks are required to 
have transverse and longitudinal impact parameters $|d_0|<0.2$~mm and $|z_0| < 1$~mm 
with respect to the primary vertex. To reduce background from heavy flavor hadrons, 
each muon is required to be isolated such that $\Sigma\pt(R<0.3)/\ptmu <0.05$, where 
only inner detector tracks with $\pt >1$~\gev\ enter the sum. Muons are required 
to have opposite electric charges. 
In cases where more 
than two muons are found to satisfy the above requirements, the pair of muons 
with the largest invariant 
mass is considered.

The dielectron and dimuon distributions are inspected for consistency with background 
predictions to ensure that the resolution and efficiency corrections were adjusted 
properly in the simulation.  Excellent agreement is found around the mass of the $Z$, 
in terms of both the peak position and width of the dilepton invariant mass 
distributions.  For the mass range $70 < m_{\ell\ell} < 110$~\gev, the number of 
events observed in data agrees to within 1\% of the background predictions for both 
the electron and muon channels.  Furthermore, the tails of the \pte\ and \ptmu\ 
distributions in the simulation are found to closely match the data.

The presence of at least one photon candidate with $\ptgam > 20$~\gev\ and 
pseudorapidity $|\eta | < 2.37$ is then necessary for the events to be kept.  Photons 
within the transition region between the barrel and the endcap calorimeters are 
excluded. Photon candidates are formed from clusters of cells in the electromagnetic 
calorimeter.  They include unconverted photons, with no associated track, and photons 
that converted to electron-positron pairs, associated to one or two tracks.  All 
photon candidates are required to satisfy the \tight\ photon 
definition~\cite{ATLASphotons}. This selection includes constraints on the energy 
leakage into the hadronic calorimeter as well as stringent requirements on the energy 
distribution in the first sampling layer of the electromagnetic calorimeter, and on 
the shower width in the second sampling layer.  The \tight\ photon definition is 
designed to increase the purity of the photon selection sample by rejecting most of 
the jet background, including jets with a leading neutral hadron (usually a $\pi^0$) 
that decays to a pair of collimated photons. To further reduce background from 
misidentified jets, photon candidates are required to be isolated by demanding that 
the sum of the transverse energies of the cells within a cone $R < 0.4$ of the photon 
be less than 10~\gev. As for the electron isolation, the core of the photon energy 
deposition is excluded and the sum is corrected for transverse shower leakage and 
pile-up.  Because no background predictions are simulated for $R(\ell,\gamma) < 0.5$, 
photons are required to be well separated from the leptons with $R(\ell,\gamma) > 
0.7$. This requirement has a negligible impact on signal efficiency.  Finally, if 
more than one photon in an event satisfies the above requirements, the one with the 
largest \pt\ is used in the search.

For the above selection criteria, the total signal acceptance times efficiency ($A\times 
\epsilon$) is $\sim 56\%$ in the $e^*$ channel for masses $m_{e^*} > 600$~\gev.  This value 
includes the acceptance of all selection cuts and the reconstruction efficiencies, and reflects 
the lepton and photon angular distributions. In comparison, $A\times \epsilon$ is $\sim 32\%$ 
for $m_{\mu^*} > 600$~\gev. The lower acceptance in the $\mu^*$ channel is due to the stringent 
selection on the muon spectrometer hits used to maximize the \ptmu\ resolution, in particular 
the limited geometrical coverage of the muon spectrometer with three layers of precision 
chambers.

\section{Background determination}

All background predictions are evaluated with simulated samples.  These include the dominant and 
irreducible \zpgam\ background, as well as \zpjet\ events where a jet is misidentified as a 
photon. The rate of jet misidentification is overestimated in the simulation so the \zpjet\ 
predictions are adjusted to data as described below.  Small contributions from \ttbar\ and 
diboson production are also present at low $m_{\ell\ell\gamma}$.  Background from multijet 
events and semileptonic decays of heavy flavor hadrons are heavily suppressed by the isolation 
requirements and are negligible in the signal region.

The \zpjet\ estimates are adjusted to data in a control region defined by 
$m_{\ell\ell\gamma} < 300$~\gev.  This region represents less than 1\% of the signal 
parameter-space for $m_\lstar \ge 200$~\gev.  The nominal strategy consists of 
counting the number of events in data in this control region and comparing it to the 
MC background predictions. The excess of background events found in the simulation is 
attributed to the mis-modeling of the rate of jets misidentified as photons, and the 
number of \zpjet\ events is scaled down accordingly.  As a result, the number of 
events in the control region is the same in the MC simulations as in data as shown in 
Table~\ref{table-control}. The \zpjet\ estimates are validated using various 
data-driven methods, notably by using misidentification rates evaluated in 
jet-enriched samples, and applying these rates to \zpjet\ data samples using an 
approach similar to the one described in Ref.~\cite{ATLASphotons}.  The main reason 
for the overestimation of the jet misidentification rate in the simulation is due to the 
mis-modeling of the jet shower shapes. A \zpjet\ 
enriched sample was used to correct the shower shapes of jets in the simulations, 
such that the efficiency for jets to pass the \tight\ photon requirement in the MC 
simulation is comparable to the rate measured in data. This correction depends 
strongly on the generator used (e.g. \pythia\ vs \alpgen) and results in a 15\% uncertainty in the \zpjet\ 
background estimate.

The largest difference between the nominal \zpjet\ background determination and the 
alternative estimates is assigned as a systematic uncertainty and dominates the total error 
in the \zpjet\ estimates presented in Table~\ref{table-control}.  The corresponding scaling 
factors applied to the \zpjet\ simulation are $0.51\pm 0.14$ and $0.61\pm 0.21$ for the 
$e^*$ and $\mu^*$ channels, respectively, i.e.~within uncertainties of one another.  
Furthermore, the ratio of the number of \zpjet\ events outside the control region to the 
number of events inside is found to be the same in the MC simulations as in the data-driven 
techniques: 0.06 for both the $e^*$ and $\mu^*$ channels. 
This finding indicates that the jet misidentification rate as a 
function of the jet \pt\ is modeled properly.

\begin{table}[!htbp]
\begin{center}
\caption{Data yields and background expectations inside ($m_{\ell\ell\gamma} < 300$~\gev)
and outside the 
$m_{\ell\ell\gamma}$ control region after adjusting the \zpjet\ 
background.
The uncertainties shown are purely statistical, except for the \zpjet\ background
for which the total uncertainty is dominated by systematic uncertainties. 
\label{table-control}}
\begin{tabular}{l||c|c|c|c||c}
\hline
\hline
Region $[$\gev$]$  & \zpgam\       & \zpjet\        & diboson        & \ttbar        & data \\
\hline
$m_{ee\gamma} < 300$  &  $306 \pm 8$  &  $138 \pm 38$  & $8.3 \pm 0.8$  & $2.4\pm 0.5$  & 455\\
$m_{ee\gamma} > 300$  &  $25 \pm 2$   &  $8.1 \pm 1.6$ & $0.8 \pm 0.2$  & $0.5\pm 0.2$  & 29\\
\hline
$m_{\mu\mu\gamma} < 300$    &  $255 \pm 8$  &  $89  \pm 31$  & $4.9 \pm 0.6$  & $0.9\pm 0.3$  & 350\\
$m_{\mu\mu\gamma} > 300$    &  $14 \pm 1$   &  $5.4 \pm 1.4$ & $0.9 \pm 0.3$  & $0.1\pm 0.1$  & 19\\
\hline
\hline
\end{tabular}
\end{center}
\end{table}

Comparisons between data and the resulting background expectations for the \ptlepton, \ptgam, 
$m_{\ell\gamma}$ and $m_{\ell\ell\gamma}$ distributions are shown in Figs.~\ref{fig:lpt} to 
\ref{fig:llg}.  No significant discrepancies are observed between data and the simulations.  In 
particular, the background prediction for the photon \pt\ shape matches the data for both the 
$e^*$ and $\mu^*$ searches, which suggests that the tuning of the jet misidentification rate for 
the \zpjet\ background is adequate.

\begin{figure}[hbpt]
\includegraphics[width=3.5truein]{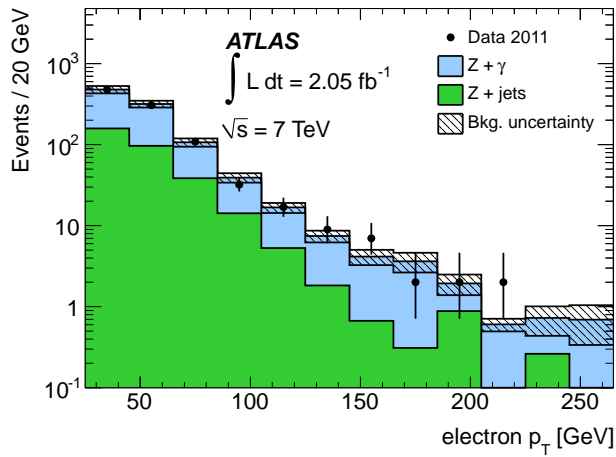}
\includegraphics[width=3.5truein]{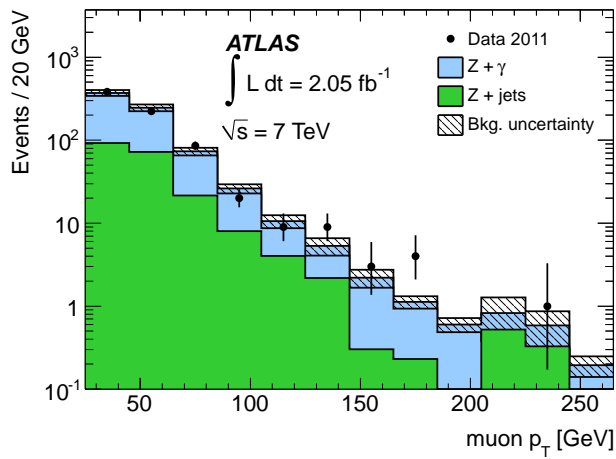}
  \caption{Lepton \pt\ distributions for the $e^*$ (top) and $\mu^*$ (bottom) channels.
The expected background uncertainties shown correspond to the sum in quadrature of the statistical uncertainties as well 
as the uncertainty in the \zpjet\ normalization measured in the control region.}
  \label{fig:lpt}
\end{figure}

\begin{figure}[hbpt]
\includegraphics[width=3.5truein]{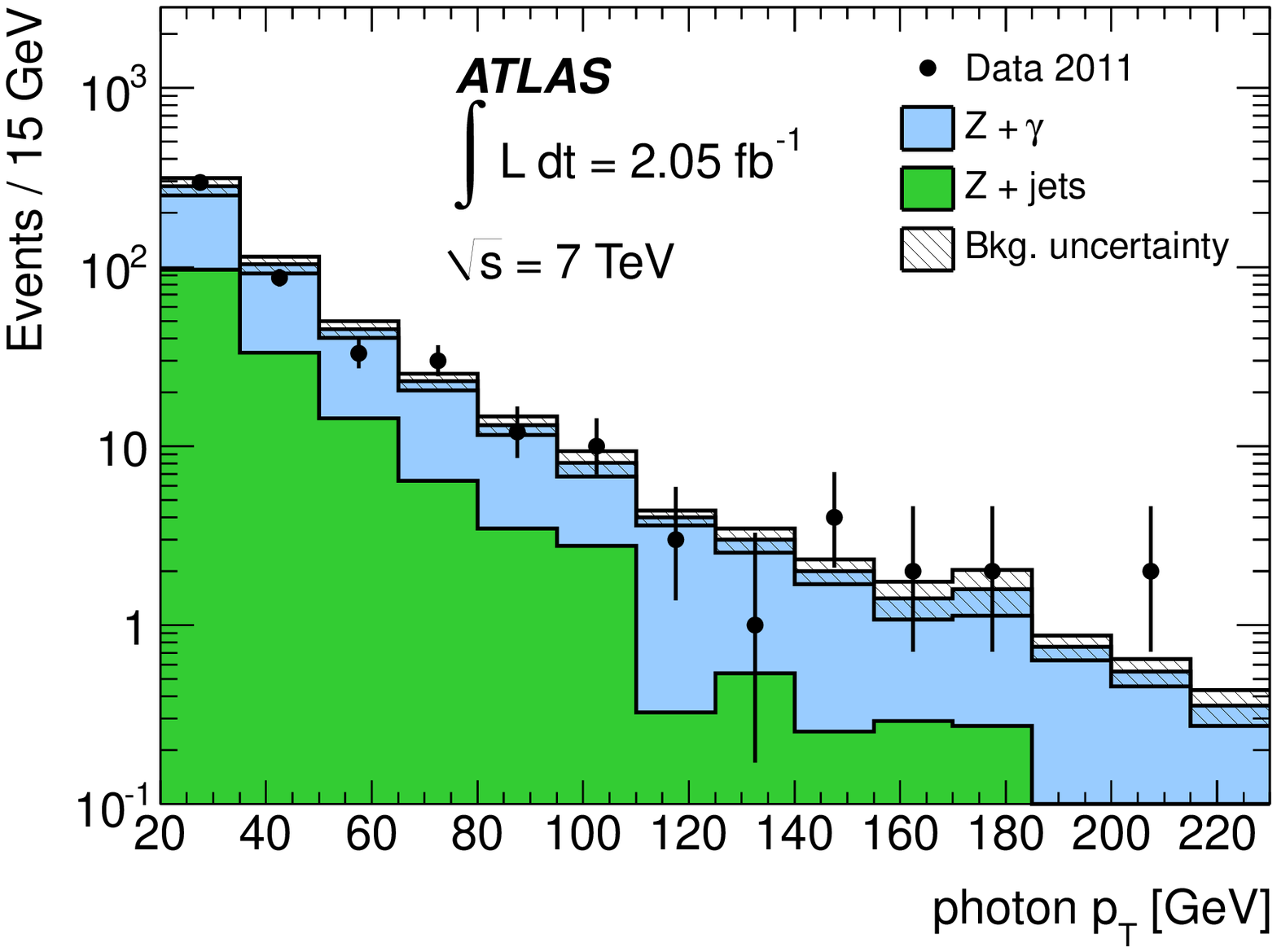}
\includegraphics[width=3.5truein]{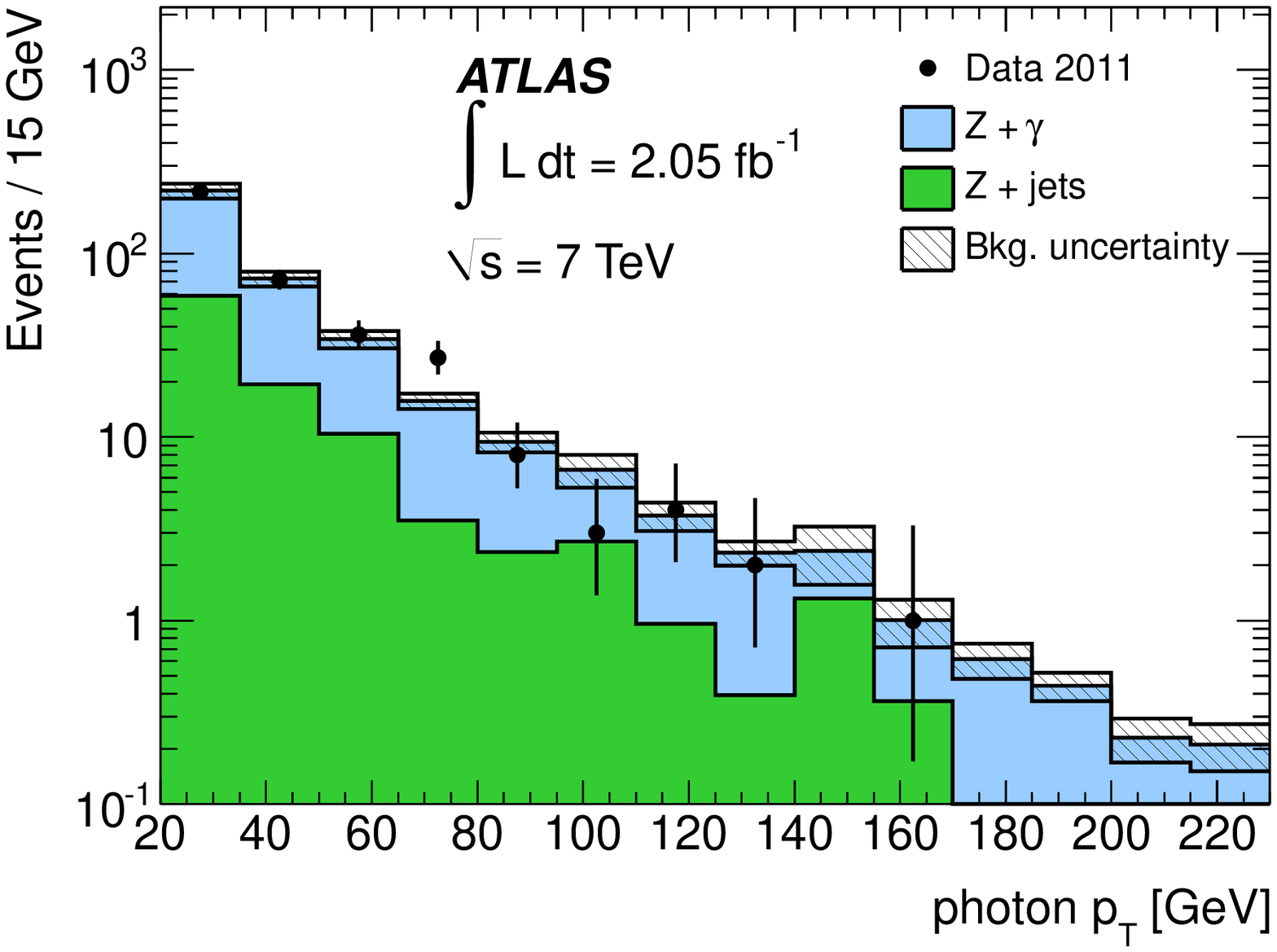}
  \caption{Photon \pt\ distributions for the $e^*$ (top) and $\mu^*$ (bottom) channels.
The expected background uncertainties shown correspond to the sum in quadrature of the statistical uncertainties as well 
as the 
uncertainty in the \zpjet\ normalization measured in the control region.}
  \label{fig:gpt}
\end{figure}

\begin{figure}[hbpt]
\includegraphics[width=3.5truein]{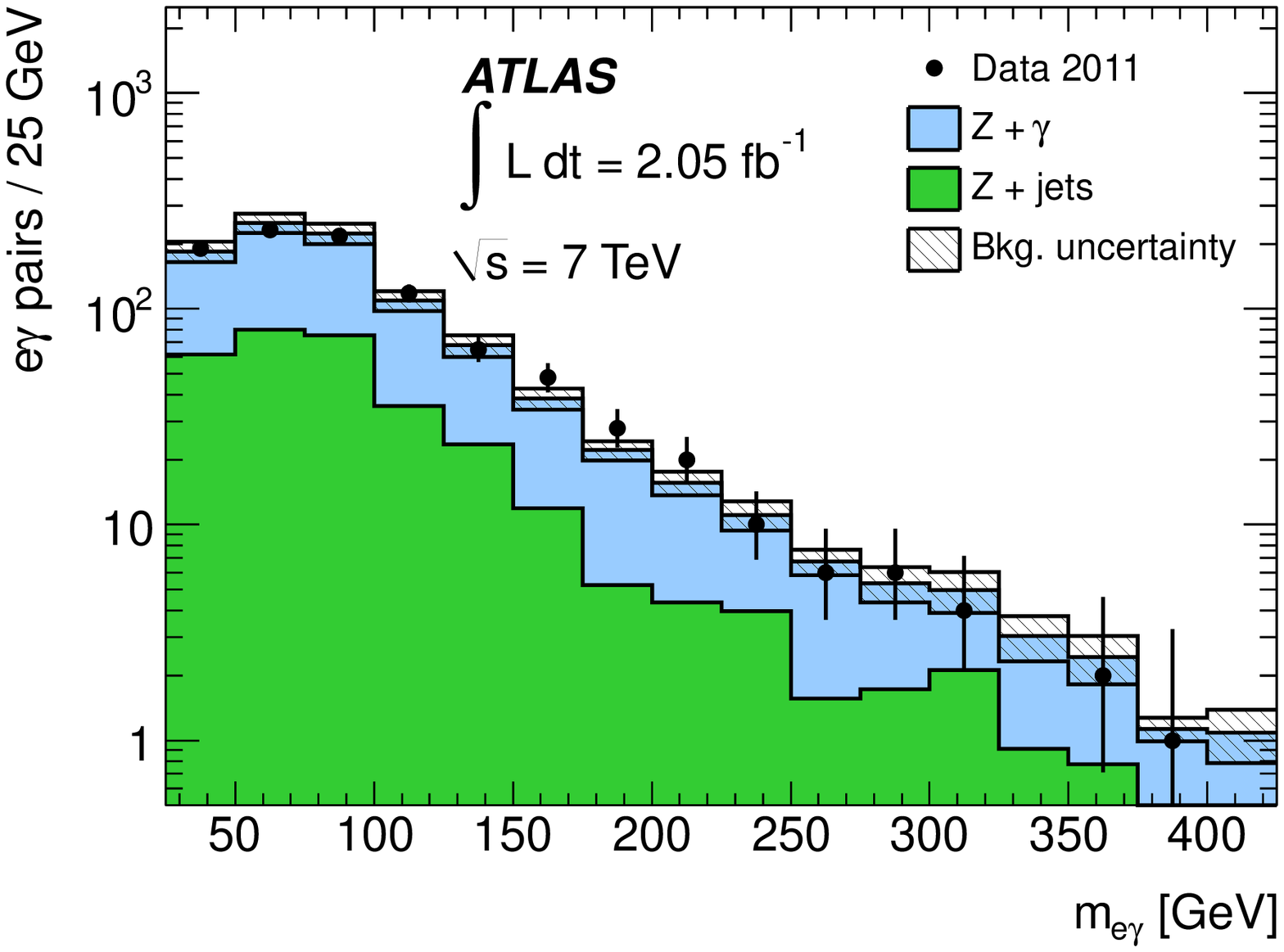}
\includegraphics[width=3.5truein]{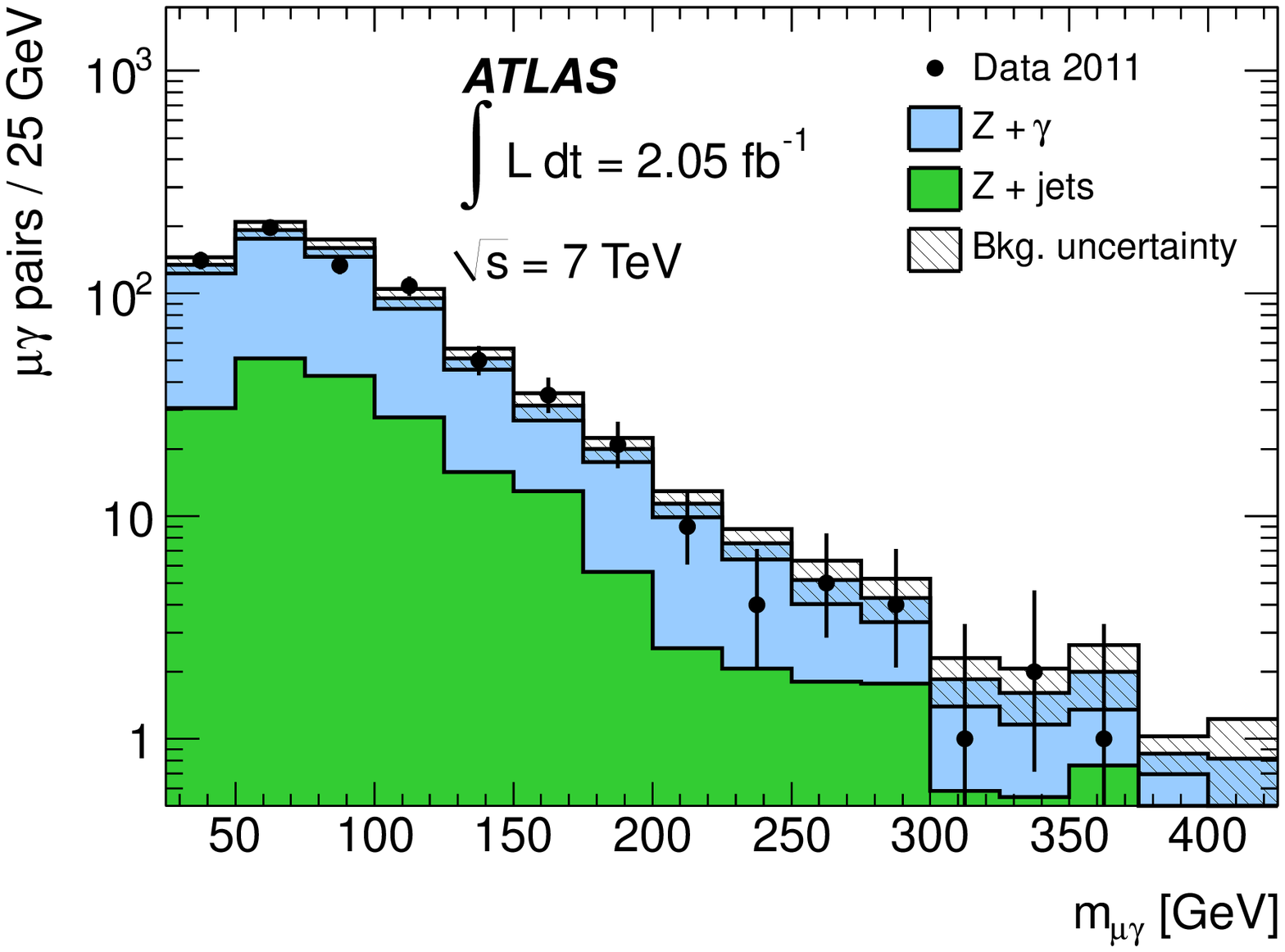}
  \caption{Distributions of the invariant mass of the $\ell\gamma$ systems for the $e^*$ (top) and $\mu^*$ (bottom) channels.
Combinations with both the leading and subleading leptons are shown.
The expected background uncertainties shown correspond to the sum in quadrature of the statistical uncertainties as well 
as the  
uncertainty in the 
\zpjet\ normalization measured in the control region.  For both channels, one event lies outside the mass range shown.}
  \label{fig:lg}
\end{figure}

\begin{figure}[hbpt]
\includegraphics[width=3.5truein]{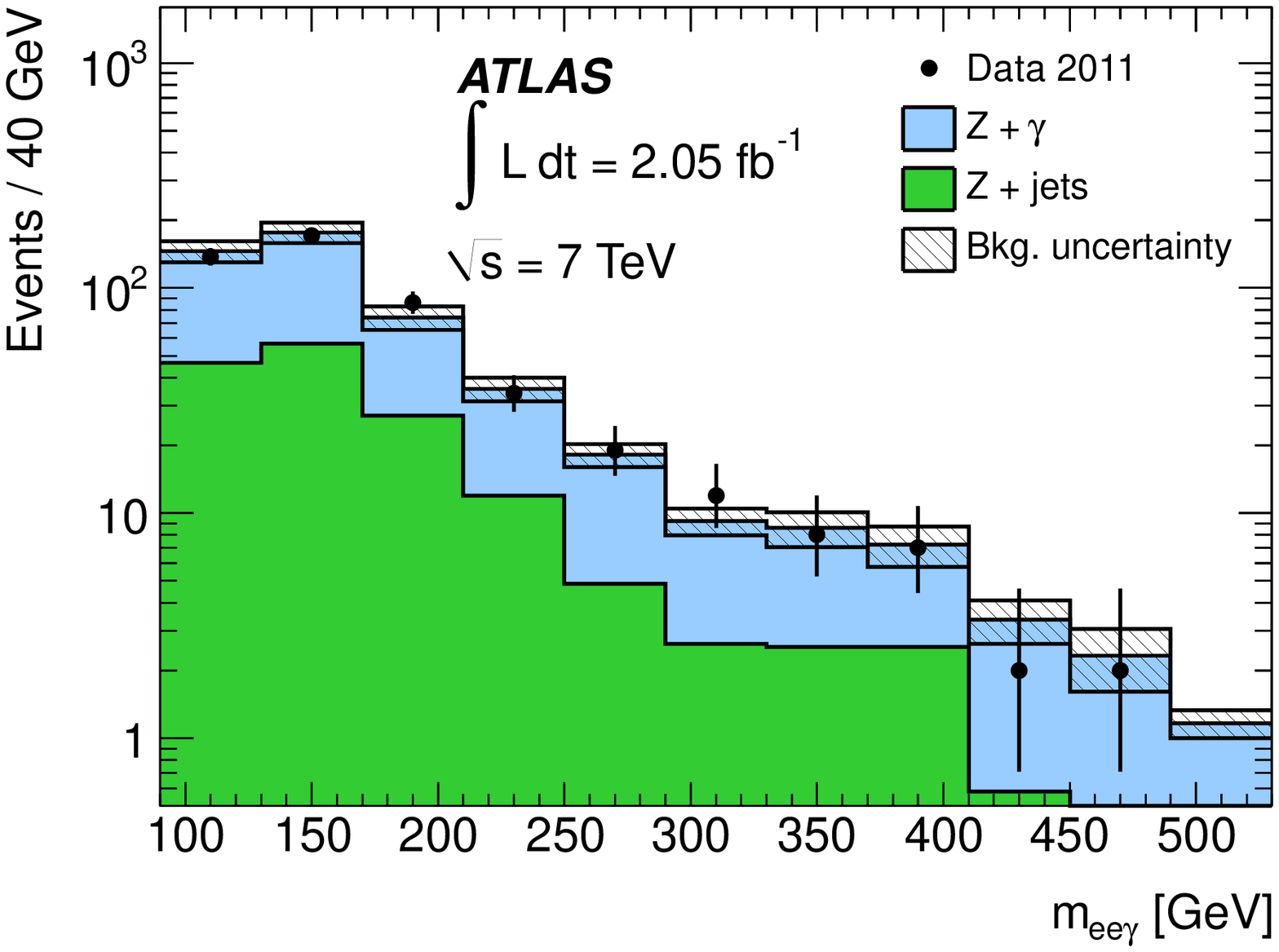}
\includegraphics[width=3.5truein]{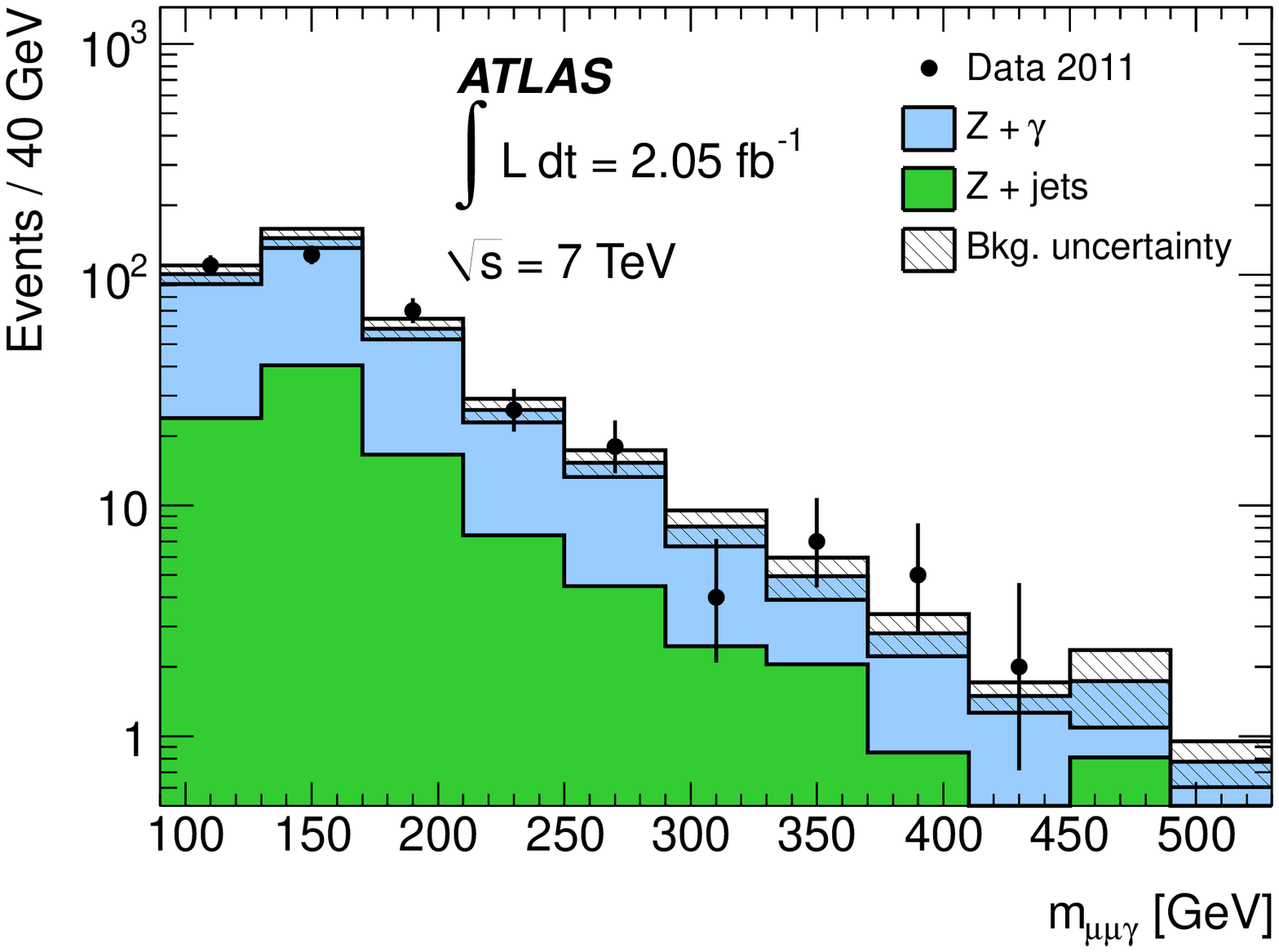}
  \caption{Distributions of the invariant mass for the $\ell\ell\gamma$ system for the $e^*$ (top) and $\mu^*$ (bottom) channels.
The expected background uncertainties shown correspond to the sum in quadrature of the statistical uncertainties as well 
as the  
uncertainty in the 
\zpjet\ normalization measured in the control region.  For both channels, one event lies outside the mass range shown.}
  \label{fig:llg}
\end{figure}

\section{Signal region optimization}

The signal search region is optimized as a function of $m_\lstar$ using simulated events by 
determining the lower bound on $m_{\ell\ell\gamma}$ that maximizes the significance defined 
as
\begin{displaymath}
S_L = \sqrt{ 2 \ln \left[ \left( 1 + S/B \right)^{S+B} e^{-S} \right] }, 
\end{displaymath}
\noindent where $S$ and $B$ are the number of signal and background events, respectively.  The 
optimum threshold value is found to be $m_{\ell\ell\gamma} = m_\lstar + 150$~\gev. Additionally, 
to improve the sensitivity particularly at low $m_\lstar$, background contributions 
from DY 
processes are suppressed further by requiring events to satisfy $m_{\ell\ell} > 110$~\gev. The 
signal efficiency for these two additional requirements is $>99$\% for $m_\lstar \ge 200$~\gev.

Because few events survive the complete set of requirements, the shape of the \zpgam\ and 
\zpjet\ backgrounds are individually fitted using an exponential function $\exp(P_0 + P_1 
\times m_{\ell\ell\gamma})$ over the mass range $250 \gev < m_{\ell\ell\gamma} < 950$~\gev. 
The sum of these two fits is then used to obtain the total background prediction for 
$m_{\ell\ell\gamma} > 350$~GeV. The resulting background estimates and data yields are shown 
in Table~\ref{tab:eyield} for the $e^*$ and $\mu^*$ searches, as well as in 
Figs.~\ref{fig:lg-final} and \ref{fig:llg-final}.

\begin{figure}[hbpt]
\includegraphics[width=3.5truein]{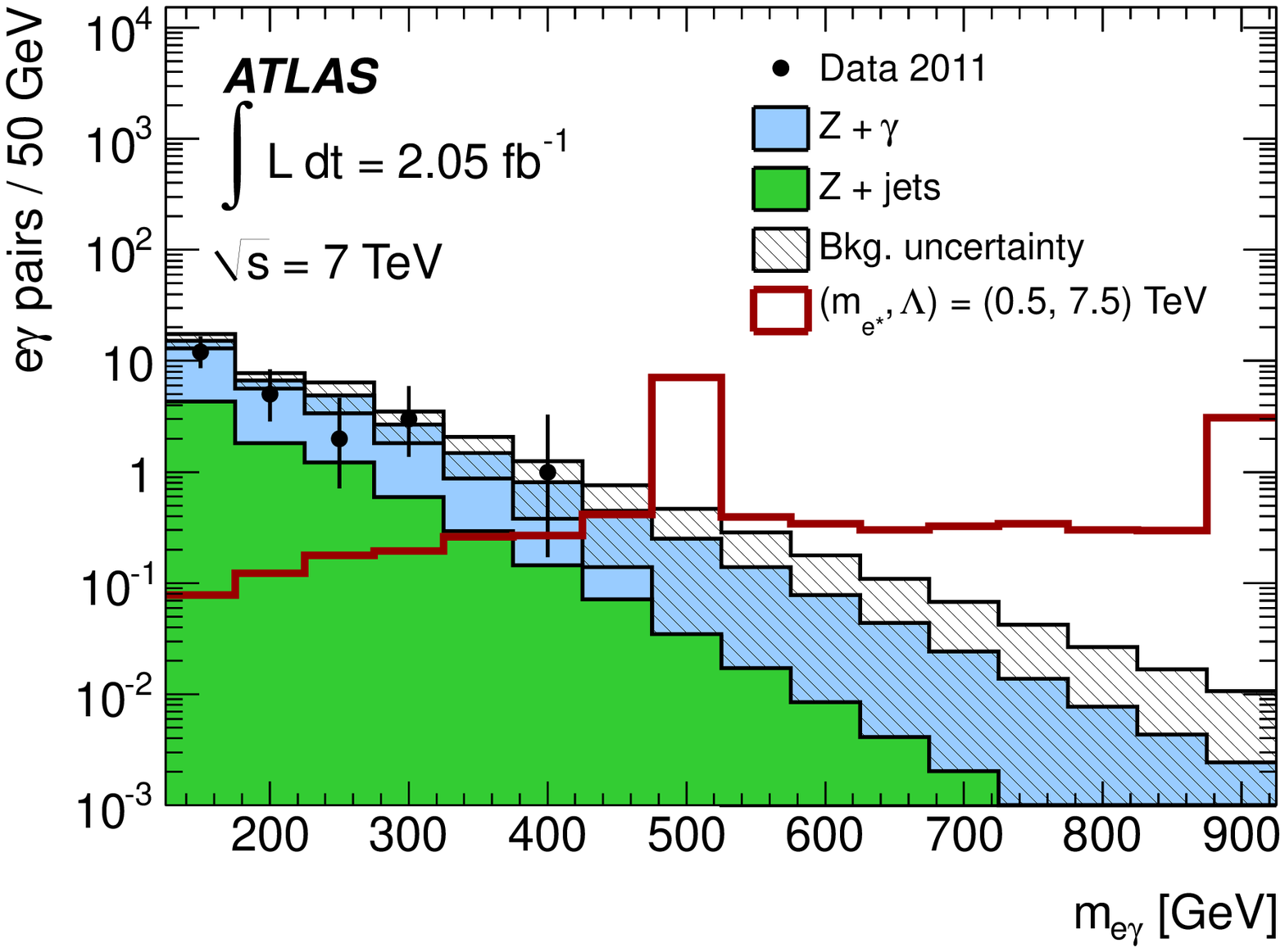}
\includegraphics[width=3.5truein]{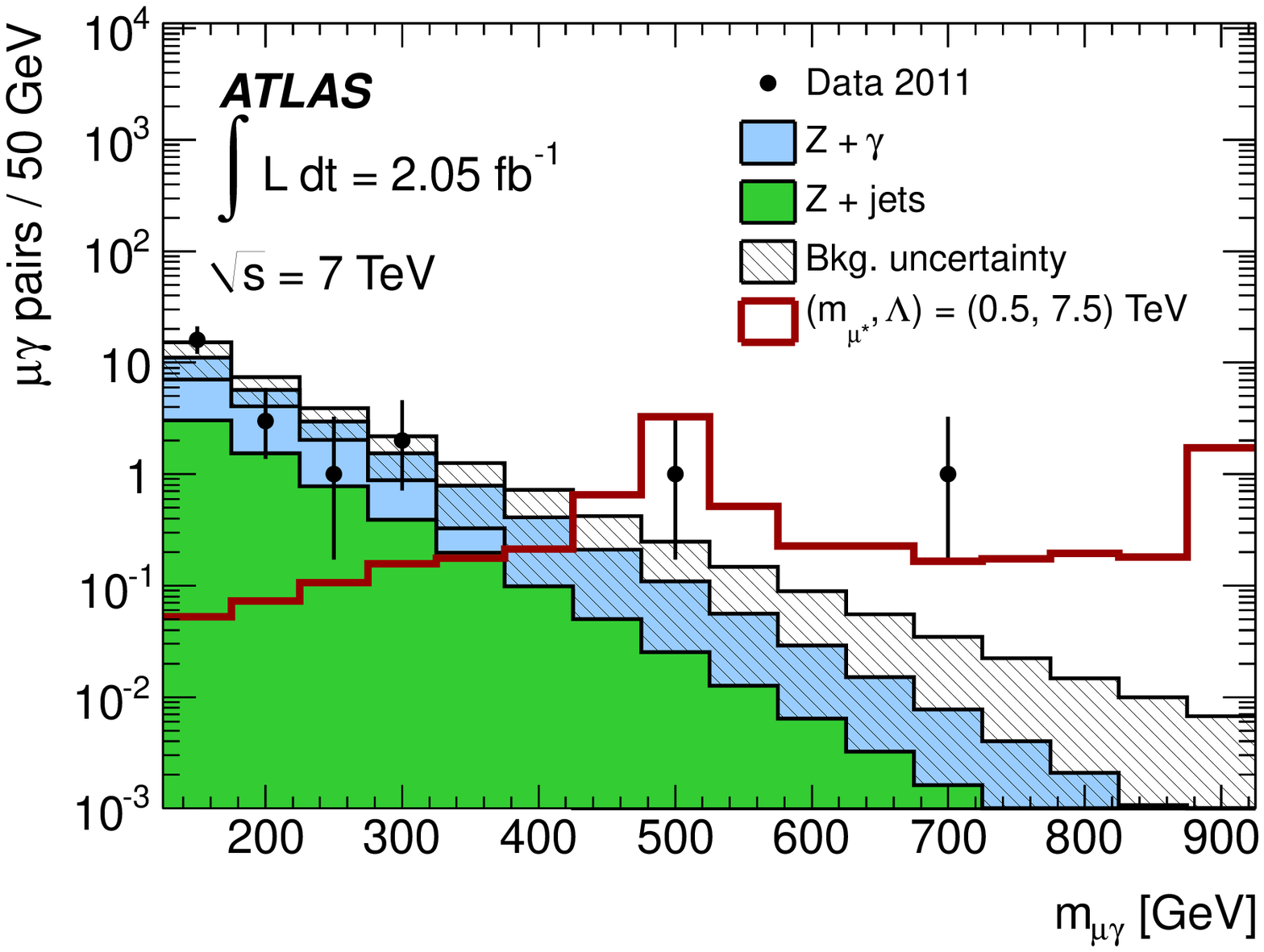}
  \caption{Distributions of the invariant mass of the $\ell\gamma$ systems for the $e^*$ (top) and $\mu^*$ 
(bottom) channels after requiring $m_{\ell\ell} > 110$~\gev.
Combinations with both the leading and subleading leptons are shown.
The expected background uncertainties shown correspond to the sum in quadrature of the statistical uncertainties as well 
as the uncertainty in the \zpjet\ normalization measured in the control region.
Note that the last bin contains the sum of all entries with $m_{\ell\gamma}> 950$~\gev.}
  \label{fig:lg-final}
\end{figure}

\begin{figure}[hbpt]
\includegraphics[width=3.5truein]{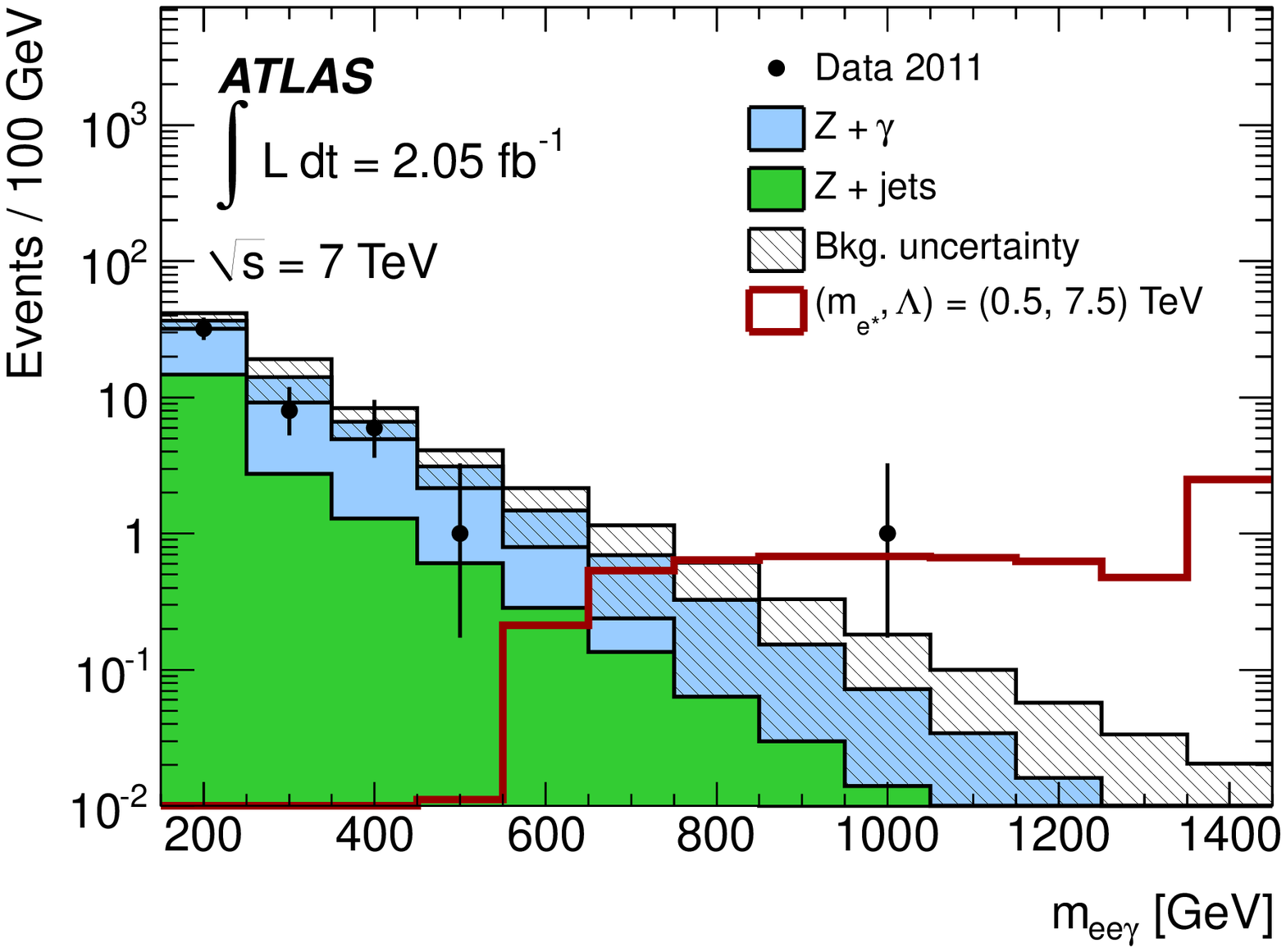}
\includegraphics[width=3.5truein]{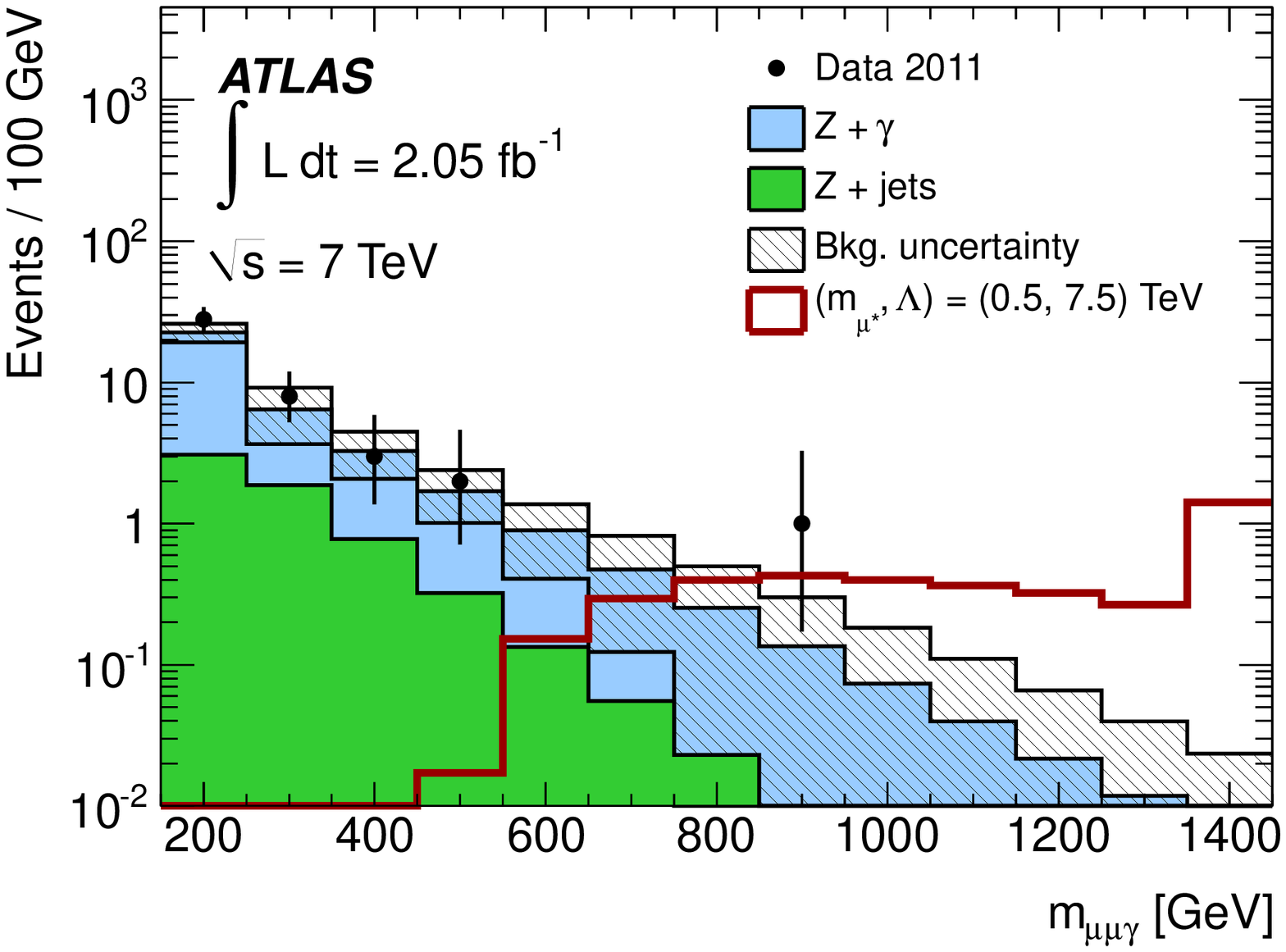}
  \caption{Distributions of the invariant mass for the $\ell\ell\gamma$ system for the $e^*$ (top) and $\mu^*$ (bottom) 
searches after requiring $m_{\ell\ell} > 110$~\gev.
The \zpjet\ and \zpgam\ backgrounds were
fitted, and the total uncertainties from the fit as well as the uncertainty in the \zpjet\
normalization measured in the control region are displayed as the shaded area. Note that the last bin contains the sum of all 
events with $m_{\ell\ell\gamma}> 1450$~\gev.}
  \label{fig:llg-final}
\end{figure}

\section{Systematic uncertainties}

In this section, the dominant systematic uncertainties in the \zpgam\ and \zpjet\ background 
predictions are first described, followed by a description of the experimental systematic 
uncertainties that affect both the background and signal yields, and by a discussion of the 
theoretical uncertainties which affect both the $e^*$ and $\mu^*$.

The dominant systematic uncertainty in the irreducible \zpgam\ background comes from the fit 
of its background shape and normalization due to the limited number of events with 
$m_{\ell\ell} > 110$~\gev.  This uncertainty increases with $m_\lstar$ from about 20\% at 
200~\gev\ to 100\% for $m_\lstar > 800$~\gev. The second largest uncertainty in the \zpgam\ 
background is of theoretical nature and arises from the NLO computations.  This uncertainty 
is obtained by varying the renormalization and factorization scales by factors of two around 
their nominal values and combining with uncertainties arising from the PDFs and values of 
the strong coupling constant $\alpha_s$. For $m_\lstar = 200$~\gev\ ($m_\lstar>800$~\gev), 
the resulting theoretical uncertainty in the number of \zpgam\ background events in our 
signal region is 7\% (10\%) for both channels.

The uncertainty in the \zpjet\ normalization is determined to be 38\% (35\%) for the 
$e^*$ ($\mu^*$) channel, which covers the range of values obtained by the different 
estimates as well as their uncertainties in the $m_{\ell\ell\gamma}<300$~GeV control region.  
Uncertainties in the \zpjet\ prediction from the shape of the fitted distribution are added in quadrature to the 
normalization uncertainty.

Experimental systematic uncertainties that affect both signal and background yields
include the uncertainty from the luminosity measurement of 3.7\% \cite{lumi2011}, and uncertainties in
particle reconstruction and identification as described below.

A 3\% systematic uncertainty is assigned to the photon efficiency.  This value is obtained by comparing the 
signal efficiency with and without photon shower shape corrections (2\%), by studying the impact of material 
mis-modeling in the inner detector (1\%) and the by determining the reconstruction efficiency for various 
pile-up conditions (1\%) \cite{diphotonExo}.

The electron trigger and reconstruction efficiency is evaluated in data and in MC 
simulations in several $\eta \times \phi$ bins to high precision.  Correction factors are 
applied to the simulations accordingly and have negligible uncertainties.  A 1\% systematic 
uncertainty in the electron efficiency at high \pt\ is assigned.  This uncertainty is 
estimated by studying the electron efficiency as a function of the calorimeter isolation 
criteria.

The calorimeter energy resolution is dominated at high \pt\ by a constant term which is 
1.1\% in the barrel and 1.8\% in the endcaps.  The simulation is adjusted to reproduce this 
resolution at high energy, and the uncertainty in this correction has a negligible effect on 
\pte\ and \ptgam.  The calorimeter energy scale is corrected by studying $J/\psi\rightarrow 
ee$ and $Z\rightarrow ee$ events.  Calibration constants are obtained for several \eta\ 
regions and deviate at most by 1.5\% of unity, and have small uncertainties.  Thus, 
uncertainties on the calorimeter energy scale and resolution result in negligible uncertainties in 
the background and signal yields.

The combined uncertainty in yields arising from the trigger and reconstruction efficiency for muons is 
estimated to increase linearly as a function of $\ptmu$ to about 1.5\% at 1~\tev. This uncertainty is 
dominated by a conservative estimate of the impact of large energy loss from muon 
bremsstrahlung in the calorimeter which can affect reconstruction in the muon 
spectrometer.  The uncertainty from the resolution due to residual misalignments in the muon 
spectrometer propagates to a change in the number of events passing the $m_{\mu\mu\gamma}$ cut, 
and affects the sensitivity of the search. The muon momentum scale is calibrated with a 
statistical precision of 0.1\% using the $Z\to\mu\mu$ mass peak.  Thus,
uncertainties on the muon momentum scale and resolution result in negligible uncertainties in 
the background and signal yields.

An additional 1\% systematic uncertainty is assigned to the $e^*$ and $\mu^*$ signal efficiencies to account for 
the fact that the dependence on $\Lambda$ is neglected in this analysis. This uncertainty is 
obtained by studying the signal $A\times \epsilon$ for various excited lepton masses and 
compositeness scales.  Theoretical uncertainties from renormalization and factorization scales 
and PDFs have negligible impact on the signal efficiency and are not included in the results 
presented below.

\section{Results}

A summary of the data yields and background expectations as a function of a lower bound on 
$m_{\ell\ell\gamma}$ is shown in Table~\ref{tab:eyield} for the $e^*$ and $\mu^*$ searches. 
The 
uncertainties displayed correspond to the sum in quadrature of the statistical and systematic 
uncertainties.  The significance of a potential excited lepton signal is estimated by a 
$p$-value, the probability of observing an outcome at least as signal-like as the one observed 
in data, assuming that a signal is absent.  The lowest $p$-values obtained are 3\% in the $e^*$ 
channel (for $m_{ee\gamma} > 950$~\gev), and 17\% in the $\mu^*$ channel (for $m_{\mu\mu\gamma} 
> 850$~\gev), which indicates that the data are consistent with the background hypothesis.

Given the absence of a signal, an upper limit on the \lstar\ cross section times branching ratio 
$\sigma B$ is determined at the 95\% C.L. using a Bayesian approach \cite{bayesianMethod} with a 
flat, positive prior on $\sigma B$.  Systematic uncertainties are incorporated in the limit 
calculation as nuisance parameters.  The limits are translated into 
bounds on the compositeness scale as a function of the mass of the excited leptons by comparing 
them with theoretical predictions of $\sigma B$ for various values of $\Lambda$.

The expected exclusion limits are determined using simulated pseudo-experiments (PE) containing 
only SM processes, by evaluating the 95\% C.L.~upper limits for each PE for 
each fixed value of $m_\lstar$. The median of the distribution of limits represents the 
expected limit. The ensemble of limits is used to find the $1\sigma$ and $2\sigma$ envelopes 
of the expected limits as a function of $m_\lstar$.

Figure~\ref{fig:limits1d} shows the 95\% C.L.~expected and observed limits on $\sigma B(\lstar 
\rightarrow \ell\gamma)$ for the $e^{*}$ and $\mu^{*}$ searches.  For $m_{\lstar}> 0.9$~\tev, the observed 
and expected limits on $\sigma B$ are 2.3~fb and 4.5~fb for the $e^{*}$ and $\mu^{*}$, respectively.  
The green and yellow bands show the expected $1\sigma$ and $2\sigma$ contours of the expected limits.  
When the expected number of background events is zero, there is an effective quantization of 
the expected limits obtained from the PE, and no downward fluctuation of the background is possible.  
These effects explain the behavior of the $1\sigma$ and $2\sigma$ contours of the expected limits for 
large \lstar\ masses. Theoretical predictions of $\sigma B$ for three different values of $\Lambda$ 
are also displayed in Fig.~\ref{fig:limits1d}, as well as the theoretical uncertainties from 
renormalization and factorization scales and PDFs for $\Lambda = 2$~TeV. These uncertainties are 
shown for illustrative purposes only and are not included in determining mass limits. The mass limits 
obtained for various $\Lambda$ values are used to produce exclusion limits on the 
$m_{\lstar}-\Lambda$ plane as shown in Fig.~\ref{fig:limits2d}. In the special case where $\Lambda = 
m_{\lstar}$, masses below 1.87~TeV and 1.75~TeV are excluded for excited electrons and muons, 
respectively.

\begin{widetext}

\begin{table}[!htbp]
\begin{center}
\caption{Data yields and background expectation as a function of a lower bound on $m_{\ell\ell\gamma} = m_\lstar + 150$~\gev.
The uncertainties represent the sum in quadrature of the statistical and
systematic uncertainties. The probability for the background only hypothesis ($p$-value) is also provided.
\label{tab:eyield}}
\begin{tabular}{c||c|c|c|c||c|c|c|c}
\hline\hline
$m_{\ell\ell\gamma}$ region   & \multicolumn{4}{c||}{$e^*$ search} & \multicolumn{4}{c}{$\mu^*$ search} \\
$[$\tev$]$    
& \hspace{0.5cm}\zpgam\hspace{0.5cm}  & \hspace{0.35cm}total bkg\hspace{0.35cm}  & \hspace{0.15cm}data\hspace{0.15cm} & \hspace{0.15cm}$p$-value\hspace{0.15cm} 
& \hspace{0.5cm}\zpgam\hspace{0.5cm}  & \hspace{0.35cm}total bkg\hspace{0.35cm}  & \hspace{0.15cm}data\hspace{0.15cm} & \hspace{0.15cm}$p$-value\hspace{0.15cm} \\
\hline
$> 0.35$     &  $10.1 \pm 1.9$   &  $11.5 \pm 2.2$  & 8    & 0.92 &  $5.2  \pm 1.4 $  &  $6.0  \pm 1.6 $ & 6    & 0.40\\
$> 0.45$     &  $4.6  \pm 1.0$   &  $5.1  \pm 1.2$  & 2    & 0.83 &  $3.1  \pm 0.8 $  &  $3.4  \pm 0.9 $ & 3    & 0.42\\
$> 0.55$     &  $2.1  \pm 0.7$   &  $2.3  \pm 0.8$  & 1    & 0.80 &  $1.8  \pm 0.6 $  &  $2.0  \pm 0.7 $ & 1    & 0.72\\
$> 0.65$     &  $0.98 \pm 0.47$  &  $1.02 \pm 0.49$ & 1    & 0.32 &  $1.09 \pm 0.49$  &  $1.14 \pm 0.51$ & 1    & 0.72\\
$> 0.75$     &  $0.45 \pm 0.29$  &  $0.46 \pm 0.30$ & 1    & 0.16 &  $0.65 \pm 0.39$  &  $0.67 \pm 0.39$ & 1    & 0.28\\
$> 0.85$     &  $0.20 \pm 0.16$  &  $0.21 \pm 0.17$ & 1    & 0.11 &  $0.39 \pm 0.29$  &  $0.39 \pm 0.29$ & 1    & 0.17\\
$> 0.95$     &  $0.09 \pm 0.09$  &  $0.10 \pm 0.09$ & 1    & 0.03 &  $0.23 \pm 0.21$  &  $0.23 \pm 0.21$ & 0    & 0.78\\
$> 1.05$     &  $0.05 \pm 0.05$  &  $0.05 \pm 0.05$ & 0    & 0.81 &  $0.14 \pm 0.14$  &  $0.14 \pm 0.14$ & 0    & 0.92\\
\hline\hline
\end{tabular}
\end{center}
\end{table}

\begin{figure}[ht!]
\includegraphics[width=3.5truein]{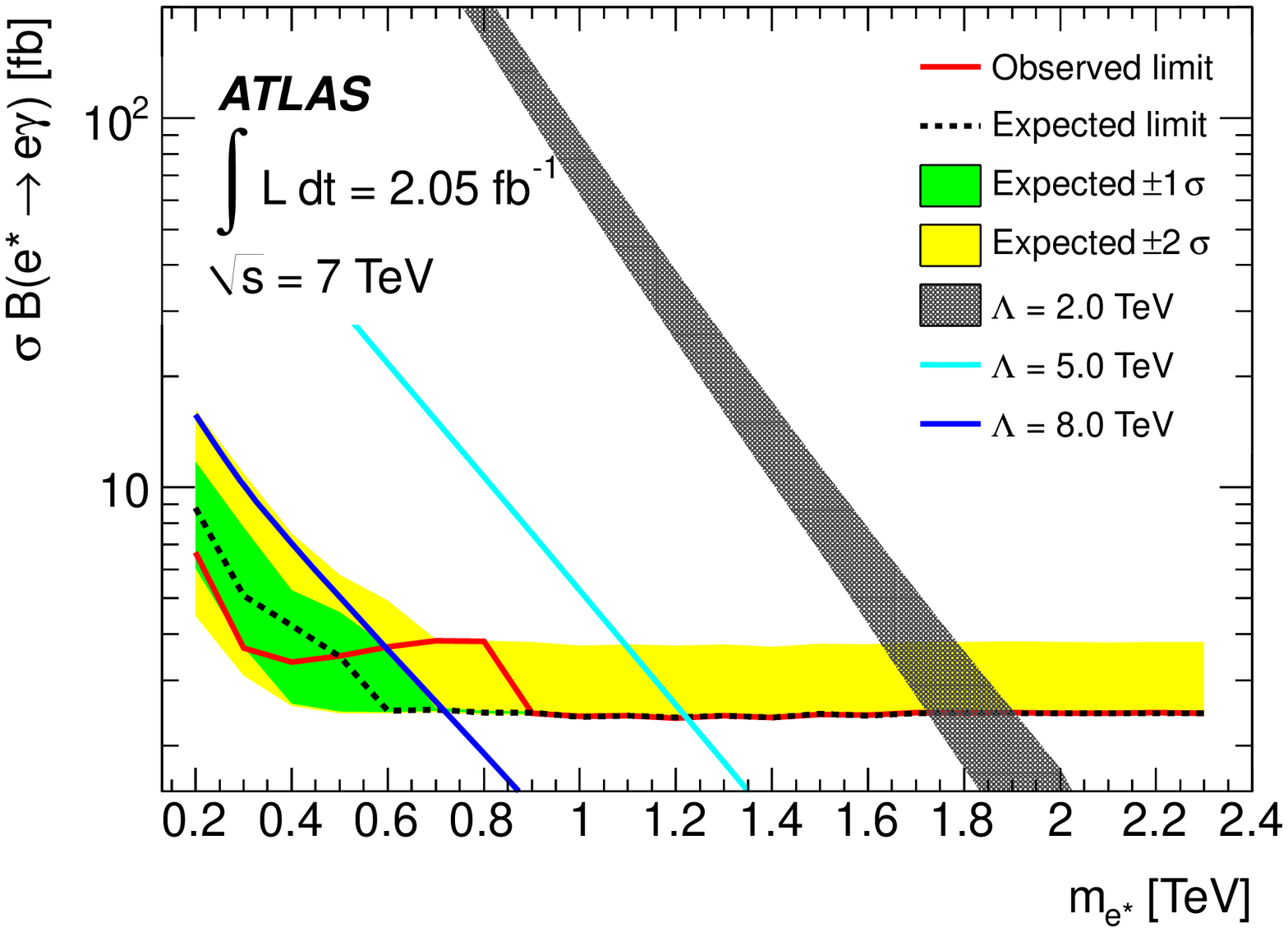}
\includegraphics[width=3.5truein]{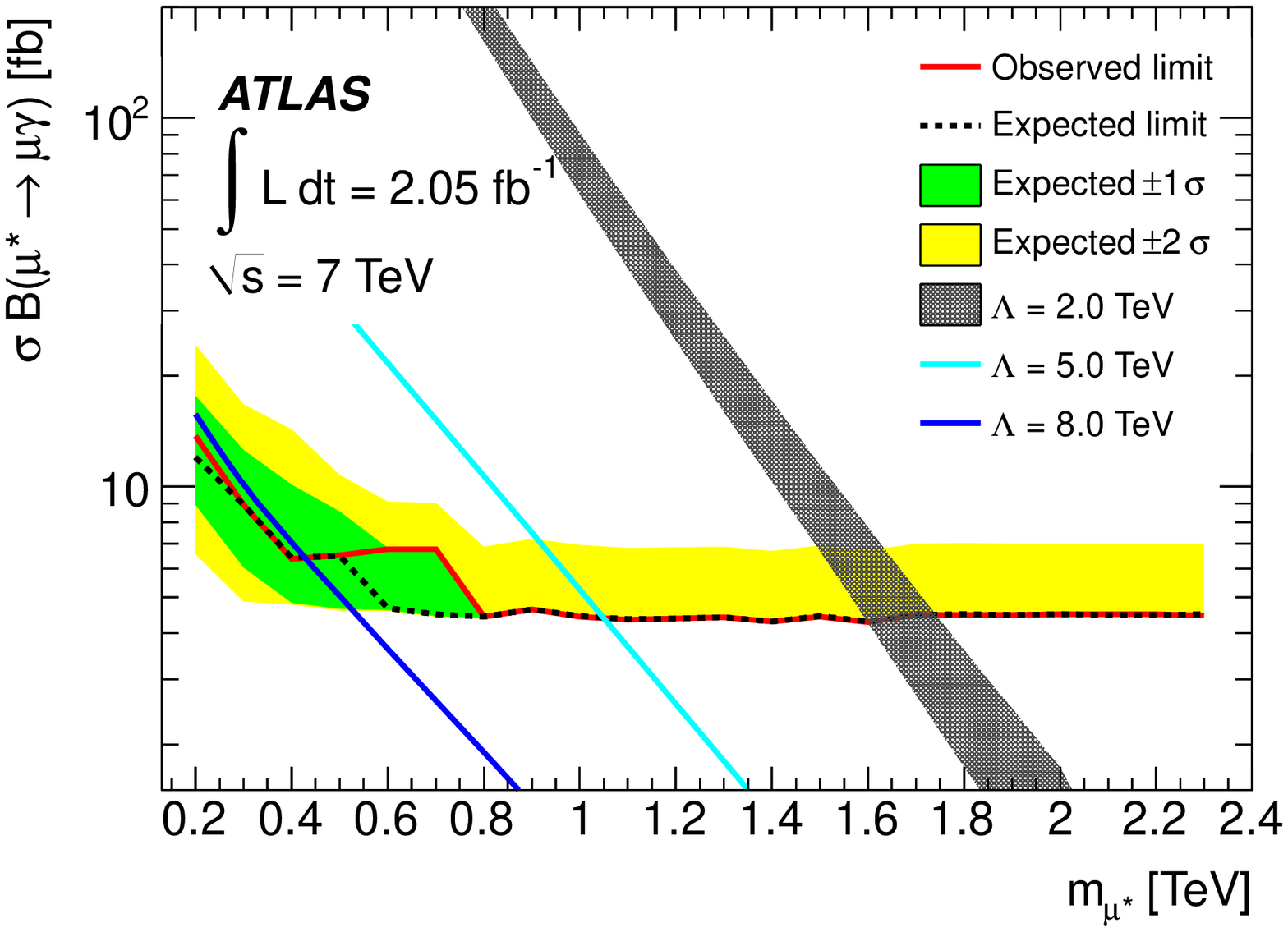}
\caption{Cross section $\times$ branching ratio limits at 95\% C.L.~as a function of $e^*$ 
and of $\mu^*$ mass.  Theoretical 
predictions for excited leptons produced for three different compositeness scales are shown, as well as the 
theoretical 
uncertainties from renormalization and factorization scales and PDFs for $\Lambda = 2$~TeV.  
For $m_{\lstar} > 0.9$~\tev, the observed limit on $\sigma B$ is 2.3~fb (4.5~fb) for $e^{*}$ ($\mu^{*}$).
\label{fig:limits1d}}
\end{figure}

\begin{figure}[hbpt]
\includegraphics[width=3.5truein]{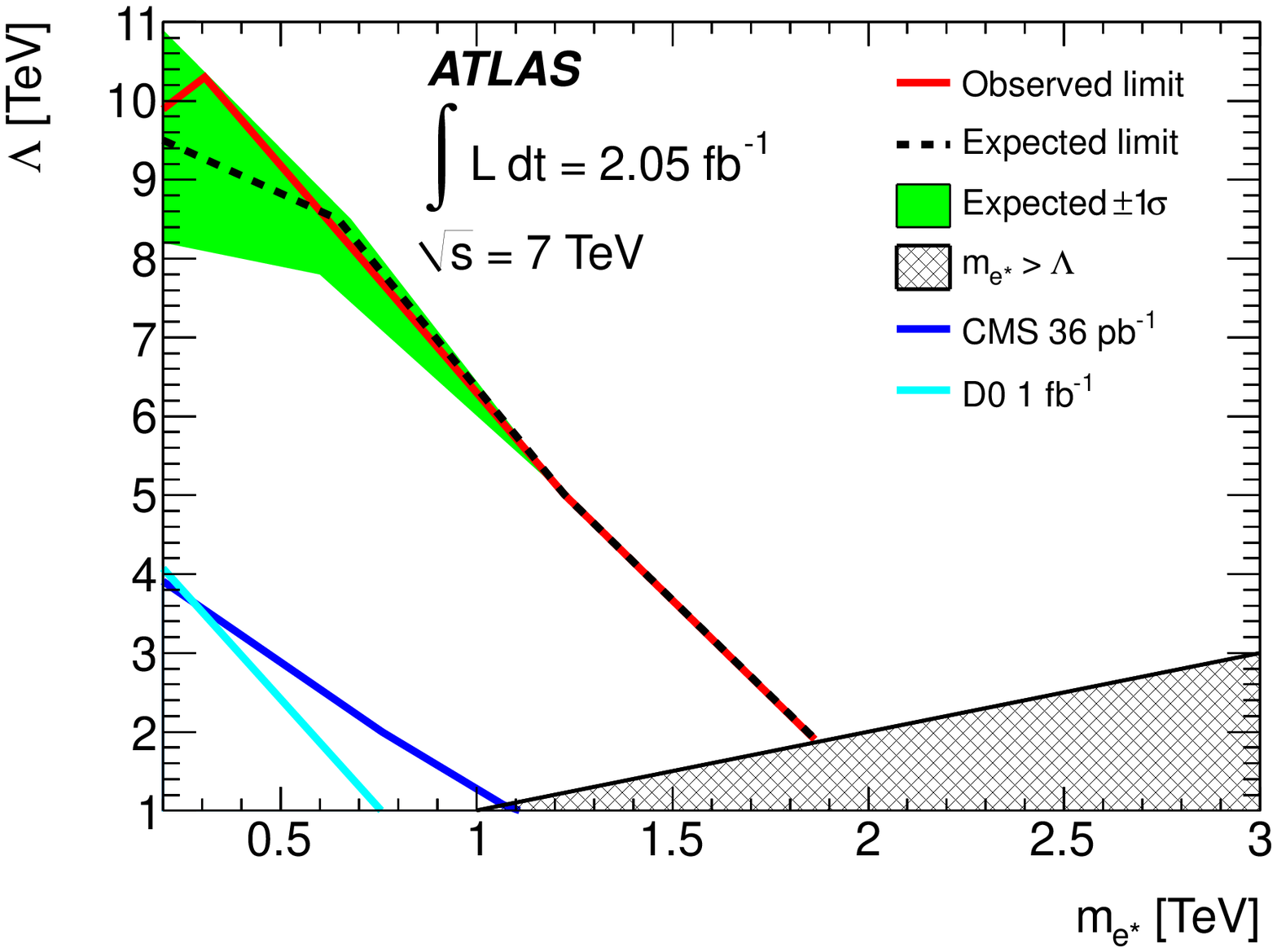}
\includegraphics[width=3.5truein]{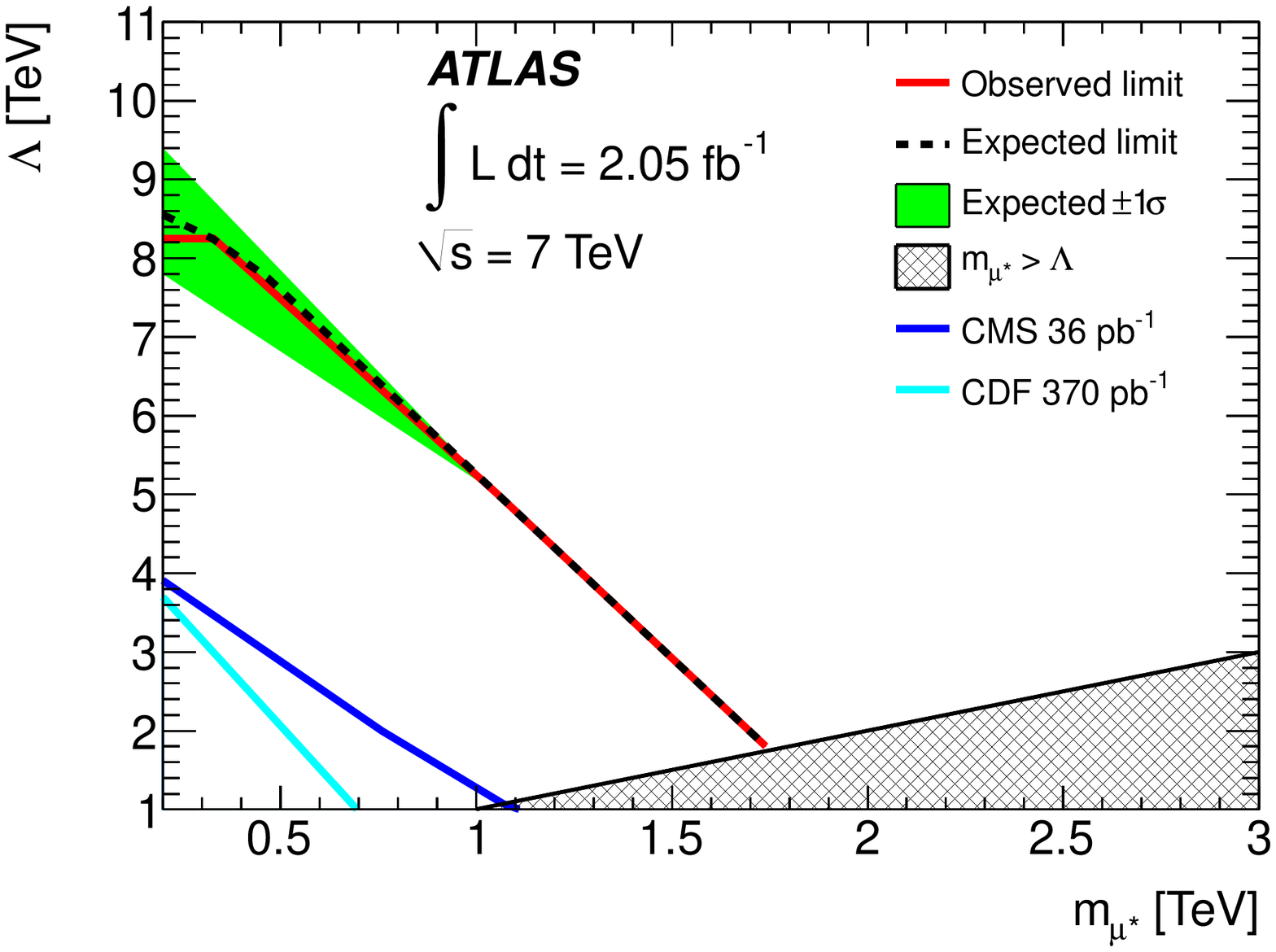}
\caption{Exclusion limits in the $m_{\lstar} - \Lambda$ parameter space for $e^*$ and $\mu^*$.
Regions to the left of the experimental limits are excluded at 95\% C.L.  No limits are set for 
the hashed region as the approximations made in the effective contact interaction model do not hold for $m_{\lstar} > 
\Lambda$.  The best limits from the Tevatron experiments as well as from the CMS experiment based on 36~pb$^{-1}$ are 
also shown.
\label{fig:limits2d}}
\end{figure}

\end{widetext}

\section{Conclusions}

The results of a search for excited electrons and muons with the ATLAS detector are reported, 
using a sample of $\sqrt{s}=7$~\tev\ $pp$ collisions corresponding to an integrated 
luminosity of 2.05 \ifb.  The observed invariant mass spectra are consistent 
with SM background expectations.  Limits are set on the cross section times branching ratio $\sigma 
B(\lstar \rightarrow \ell\gamma)$ at 95\% C.L.  For $m_{\lstar} > 0.9$~\tev, the observed upper 
limits on $\sigma B$ are 2.3~fb and 4.5~fb for the $e^{*}$ and $\mu^{*}$ channels, respectively.  The 
limits are translated into bounds on the compositeness scale $\Lambda$ as a function of the mass 
of the excited leptons. In the special case where $\Lambda = m_{\lstar}$, masses below 1.87~TeV 
and 1.75~TeV are excluded for $e^*$ and $\mu^*$, respectively. These limits are the most stringent 
bounds to date on excited leptons for the parameter-space region with $m_{\lstar} \ge 200$~\gev.


\section{Acknowledgements}

We thank CERN for the very successful operation of the LHC, as well as the
support staff from our institutions without whom ATLAS could not be
operated efficiently.

We acknowledge the support of ANPCyT, Argentina; YerPhI, Armenia; ARC,
Australia; BMWF, Austria; ANAS, Azerbaijan; SSTC, Belarus; CNPq and FAPESP,
Brazil; NSERC, NRC and CFI, Canada; CERN; CONICYT, Chile; CAS, MOST and
NSFC, China; COLCIENCIAS, Colombia; MSMT CR, MPO CR and VSC CR, Czech
Republic; DNRF, DNSRC and Lundbeck Foundation, Denmark; ARTEMIS, European
Union; IN2P3-CNRS, CEA-DSM/IRFU, France; GNAS, Georgia; BMBF, DFG, HGF, MPG
and AvH Foundation, Germany; GSRT, Greece; ISF, MINERVA, GIF, DIP and
Benoziyo Center, Israel; INFN, Italy; MEXT and JSPS, Japan; CNRST, Morocco;
FOM and NWO, Netherlands; RCN, Norway; MNiSW, Poland; GRICES and FCT,
Portugal; MERYS (MECTS), Romania; MES of Russia and ROSATOM, Russian
Federation; JINR; MSTD, Serbia; MSSR, Slovakia; ARRS and MVZT, Slovenia;
DST/NRF, South Africa; MICINN, Spain; SRC and Wallenberg Foundation,
Sweden; SER, SNSF and Cantons of Bern and Geneva, Switzerland; NSC, Taiwan;
TAEK, Turkey; STFC, the Royal Society and Leverhulme Trust, United Kingdom;
DOE and NSF, United States of America.

The crucial computing support from all WLCG partners is acknowledged
gratefully, in particular from CERN and the ATLAS Tier-1 facilities at
TRIUMF (Canada), NDGF (Denmark, Norway, Sweden), CC-IN2P3 (France),
KIT/GridKA (Germany), INFN-CNAF (Italy), NL-T1 (Netherlands), PIC (Spain),
ASGC (Taiwan), RAL (UK) and BNL (USA) and in the Tier-2 facilities
worldwide.

 
\bibliography{lstar_prd2011}{}
\onecolumngrid
\clearpage
\onecolumngrid
\include{atlas_authlist}

\end{document}

%% file: atlas_authlist.tex
\begin{flushleft}
{\Large The ATLAS Collaboration}

\bigskip

G.~Aad$^{\rm 48}$,
B.~Abbott$^{\rm 110}$,
J.~Abdallah$^{\rm 11}$,
A.A.~Abdelalim$^{\rm 49}$,
A.~Abdesselam$^{\rm 117}$,
O.~Abdinov$^{\rm 10}$,
B.~Abi$^{\rm 111}$,
M.~Abolins$^{\rm 87}$,
O.S.~AbouZeid$^{\rm 157}$,
H.~Abramowicz$^{\rm 152}$
% ...
\bigskip

% ...
$^{10}$ Institute of Physics, Azerbaijan Academy of Sciences, Baku, Azerbaijan\\
$^{11}$ Institut de F\'isica d'Altes Energies and Departament de F\'isica de la Universitat Aut\`onoma  de Barcelona and ICREA, Barcelona, Spain\\
% ...
$^{48}$ Fakult\"{a}t f\"{u}r Mathematik und Physik, Albert-Ludwigs-Universit\"{a}t, Freiburg i.Br., Germany\\
$^{49}$ Section de Physique, Universit\'e de Gen\`eve, Geneva, Switzerland\\
% ...
$^{87}$ Department of Physics and Astronomy, Michigan State University, East Lansing MI, United States of America\\
% ...
$^{110}$ Homer L. Dodge Department of Physics and Astronomy, University of Oklahoma, Norman OK, United States of America\\
$^{111}$ Department of Physics, Oklahoma State University, Stillwater OK, United States of America\\
% ...
$^{117}$ Department of Physics, Oxford University, Oxford, United Kingdom\\
% ...
$^{152}$ Raymond and Beverly Sackler School of Physics and Astronomy, Tel Aviv University, Tel Aviv, Israel\\
% ...
$^{157}$ Department of Physics, University of Toronto, Toronto ON, Canada\\
% ...
\end{flushleft}


%% file: lstar_prd2011.bbl
\begin{thebibliography}{37}
\bibitem[{coo()}]{coordinates}
\bibinfo{note}{ATLAS uses a right-handed coordinate system with the $z$-axis
  along the beam pipe. The $x$-axis points to the center of the LHC ring, and
  the $y$-axis points upward. Cylindrical coordinates $(r,\phi)$ are used in
  the transverse plane, $\phi$ being the azimuthal angle. The pseudorapidity is
  defined in terms of the polar angle $\theta$ as $\eta=-\ln\tan(\theta/2)$.}
\end{thebibliography}
